# Automatically Generate Steganographic Text Based on Markov Model and Huffman Coding

Yang Zhongliang[*,1,2], Jin Shuyu[3], Huang Yongfeng[1,2] , Zhang Yujin[1] and Li Hui[3]

[1]Department of Electronic Engineering, Tsinghua University, Beijing, 100084, China (Phone: 188-1153-6956; e-mail: yangzl15@mails.tsinghua.edu.cn).
[2]Tsinghua National Laboratory of Information Science and Technology, Beijing 10084, China
[3]School of Information Science and Engineering, Shenyang University of Technology, 110870, Liao Ning, China

**ABSTRACT**

Steganography, as one of the three basic information security systems, has long played an important role in safeguarding the privacy and confidentiality of data in cyberspace. Text is the most widely used information carrier in people's daily life, using text as carrier for information hiding has broad research prospects. However, due to the high coding degree and less information redundancy in text, it has been an extremely challenging problem to hide information in it for a long time. In this paper, we propose a steganography method which can automatically generate steganographic text based on Markov chain model and Huffman coding. It can automatically generate fluent text carrier in terms of secret information which need to be embedded. The proposed model can learn from a large number of samples written by people and obtain a good estimate of the statistical language model. We evaluated the proposed model from several perspectives. Experimental results show that the performance of the proposed model is superior to all the previous related methods in terms of information imperceptibility and information hidden capacity.

## 1. INTRODUCTION

In Shannon's monograph on information security [1], he summarized the three most basic information security systems: encryption system, privacy system, and concealment system. The encryption system, which is highlighted by Shannon, encrypts secret messages by using special coding methods. It ensures the security of information by making the message indecipherable. The privacy system is mainly to restrict access to information, so that only authorized users can access important information. Unauthorized users cannot access it by any means under any circumstances. However, while ensuring information security, these two systems also expose the existence and importance of secret information, making it more vulnerable to attacks such as interception and cracking [2]. The concealment system is very different from these two secrecy systems, it uses various carriers to embed secret information and then transmit through public channels, hide the existence of secret information , which to achieve the purpose of not being easily suspected and attacked [3]. Due to its extremely strong information concealment, steganographic system plays an important role in protecting trade secrets, military security and even national defense security.

Steganography is the key technology in a concealment system, it shares many common features with the related but fundamentally quite different from data-hiding field called watermarking [4,5]. Although both steganography and digital watermarking techniques hide information in the carrier, the primary goal of steganography is to hide the existence of information. However, for digital watermarking, the primary goal is to resist modification. Secondly, we usually hope that the secret information embedded in the concealment system as much as possible. However, for digital watermarking technology, the amount of information embedded is generally small. In addition, messages embedded in general digital watermarking systems are well designed, but the messages embedded in the concealment system are irregular.

There are various media forms of carrier that can be used for information hiding, including image [6,7], audio [8,9], text [10−17] and so on [18]. Text is the most widely used information carrier in people's daily life. Therefore, using text as a carrier to realize information hiding has great research value and practical significance. However, compared with image and audio, texts have a higher degree of information coding, resulting in less redundant information, which makes it quite challenging to use text as a carrier for information hiding [7]. For the above reasons, text steganography has attracted a large number of researchers' interests. In recent years, more and more text based on information hiding methods have emerged [11,17].

Previous works on text steganography can be divided into two big families: format based method [19] and content based method [20]. Text format based methods usually treat text as a specially coded image, they usually use the format information of the documents in terms of the organizational structure and layout of the document content to hide secret information, such as the paragraph format and the font format. For example, some of the previous works show that they can conceal information by adjusting the format of the text, like inter-character space [21], word-shifting [22], character-coding [23], etc. This type of method usually has strong visual concealment. The biggest drawback of this type of method is the poor anti-interference ability and leading to the hidden information being easily destroyed.

Content based method, also called natural language information hiding [24], is mainly based on linguistic and statistical knowledge, using Natural Language Processing(NLP) technology to make modifications to the existing normal texts in terms of vocabulary, syntax, semantics and so on, and try to keep the text of local and global semantic invariant, grammatically correct, syntactic structure reasonable to achieve information hiding. They implement information hiding by replacing some of the words in the sentences [25], or by changing their syntactical structure [26]. Such methods need to ensure that the modified text satisfies the requirements of semantic correctness and grammatical rationality. But generally, the information hiding efficiency of this method is very low.

In content-based text steganography, there are plenty of works that utilized text generation algorithms to conduct information hiding [17,27,28]. Through some natural language processing methods, they automatically generate a piece of text, and finally achieve information hiding by properly encoding the words during text generation. This type of method usually has a high hidden capacity and is therefore considered a very promising research direction in the field of text steganography. The biggest challenge with this type of method is that they need to ensure the quality of the generated text with hidden information inside is high enough.

In this paper, we propose a text automatic generation steganography method based on Markov chain model and Huffman coding. It can automatically generate fluent text carrier in terms of secret information which need to be embedded. During the process of text generation, on the one hand, we try to keep the statistical distribution of generated text similar to that of the training text. On the other hand, in the information embedding process, we dynamically encode each word according to the differences in their conditional probability distributions. By adjusting the encoding method, we can adjust the embedding rate of secret message, so that we can ensure the concealment and hidden capacity be optimized at the same time through fine control. Compared to previous works, the quality of steganographic text generated by the proposed model has increased greatly.

In the remainder of this paper, Section II introduces related steganography method based on automatic text generation. A detailed explanation of the proposed model and algorithm details of information hiding and extracting are elaborated in Section III. The following part, Section IV, presents the experimental evaluation results and gives a comprehensive discussion. Finally, conclusions are drawn in Section V.

## 2. RELATED WORK

Compared with other types text steganography methods, the methods based on automatic text generation are characterized by the fact that they do not need to be given carrier texts in advance. Instead, they can automatically generate a textual carrier based on secret information. Since these kind of method can usually achieve a high hidden capacity, it is considered to be a very promising research topic in the current steganography field.

In the early stage, in order to ensure that the generated text is consistent with the training sample in the probability distribution of the characters, Wayner's algorithm [29] mimics statistical characteristics of a normal file, then generates character sequences having similar statistic profile with the original file. By this method, it is resilient against statistical attacks but the texts they generated are meaningless. In addition, Chapman et al. [30] tried to use syntactic template or syntax structure tree to generate texts, they expected the generated texts could conform these syntactic rules. Obviously, the texts generated by this method have a very simple pattern, and they generally don't look very smooth.

Therefore, a lot of researchers combined text steganography with statistical natural language processing, and a large number of natural language processing techniques have been used to automatically generate steganographic text [10,11,17,28]. Since the Markov chain model is very suitable for modeling natural text, in recent years, a large number of works using the Markov chain model for automatic generation of steganographic text have appeared [10,11]. Most of these works use the Markov chain model to calculate the number of common occurrences of each phrase in the training set and obtain the transition probability. Then the transition probability can be used to encode the words and achieve the purpose of embedding secret information in the text generation process. This kind of method greatly improved the quality of the generated texts compared to the previous methods.

Steganography based on automatic text generation can usually be divided into two steps, one is automatic text generation and the other is secret information embedding. To generate high concealment steganographic text, we need to ensure that both steps are consistent with the statistical distribution of the training samples. However, the previous work usually only focuses on the first step, which is, the automatic text generation process ignores the second step. The result is that the generated steganographic text is of poor quality and can be easily detected. For example, Dai et al. [10] proposed a text steganography system based on Markov Chain source model and DES algorithm. However, in the process of generating steganographic texts, they ignored the difference in the transition probability of each word and fixed-length encoding each candidate word, resulting in poor quality of the generated steganographic text. Moraldo et al. [11] also proposed a method for automatic generation of steganographic texts based on Markov model, but their model mainly focus on how to ensure that each sentence generated is embedded with a fixed number of secret bits. Their model also ignored the difference in the transition probability of each word during the iteration, so the quality of the generated text cannot reach a satisfactory effect.

Considering the difficulty of automatically generating high quality steganographic text, some researchers began to try to generate text in a special format to achieve information hiding [17,28]. The advantage of these special-genre texts is that they have their own specific structure and pattern. It is easy for them to learn the rules of writing, and then makes the poetries they generated looks real enough. Desoky [24] exploited many special text forms, such as notes, jokes, chess, etc. Luo [17] developed a Ci-Based Steganography Methodology (Cistega), which uses Markov model to generate Ci-poetry, a traditional Chinese literature style. During the generating process, they choose words which meet rhythm rules from the markov transfer matrix and put them into StackList, then select a specific word according to the bit stream of secret message. But we have to realize that Chinese poetry, after all, is a kind of special-genre text, which is not often used in daily life and is also hard for most people to understand.

In this paper, combined Markov model and Huffman coding, we propose a new method to automatic generate steganographic text. It can automatically generate fluent steganographic text in terms of secret information that need to be embedded. The proposed model considers these two steps at the same time. Firstly, the Markov chain model is used to ensure that the automatic text generation process conforms to the statistical language distribution of the training samples. In addition, in the information embedding stage, each word is dynamically coded using the Huffman tree to obey the conditional probability distribution of each word. In this way, the quality of steganographic text generated by using the proposed model has increased greatly, and significantly improves information imperceptibility and information hidden capacity of the whole concealment system.

## 3. OUR METHOD

### 3.1 Automatic Text Generation Based on Markov Chain Model

In the field of statistical natural language processing, they usually use statistical language model to model a sentence. A language model is a probability distribution over sequences of words, it can be expressed by the following formula:

$$
\begin{aligned}
p(S) &= p(w_1, w_2, w_3, \dots, w_n) \\
&= p(w_1)p(w_2|w_1) \dots p(w_n|w_1, w_2, \dots, w_{n-1})
\end{aligned} \quad (1)
$$

where $S$ denotes the whole sentence with a length of $n$ and $w_i$ denotes the $i$-th word in it. $p(S)$ assigns the probability to the whole sequence. It is actually composed of the product of $n$ conditional probabilities, each of the conditional probability calculates the probability distribution of the $n$-th word when the first $n-1$ words are given, that is $p(w_n|w_1, w_2, \dots, w_{n-1})$. Therefore, in order to automatically generate high quality texts, we need to obtain a good estimate of the statistical language model of the training sample set.

In probability theory, a Markov chain is a stochastic model describing a sequence of possible events in which the probability of each event only depends on the state attained in the previous event. The Markov chain model is suitable for modeling time series signals. For instance, suppose there is a value space $\chi = \{x_1, x_2, x_3, \dots, x_m\}$, and $Q = \{q_1, q_2, q_3, \dots, q_n\}$ is a stochastic variable sequence, whose values are sampled from $\chi$. For the convenience of the following description, we will record the value of $t$-th state as $x^t$, that is $q_t = x^t, x^t \in \chi$. If we think that the value of the state at each moment in the sequence is related to the state of all previous moments, that is $p(q_t|q_1, q_2, \dots, q_{t-1})$, then the Markov chain model can be expressed as follows:

$$
\begin{aligned}
&P(q_t = x^t) \\
&= f\big(P(q_{t-1} = x^{t-1}), P(q_{t-2} = x^{t-2}), \dots, P(q_1 = x^1)\big), \\
&\text{s. t.} \quad \sum_{i=1}^{m} P(q_t = x_i) = 1, \forall x_i \in \chi
\end{aligned} \quad (2)
$$

where $f$ is the probability transfer function. Then the probability of the whole sequence can be expressed as follows:

$$
\begin{aligned}
p(Q) &= p(q_1, q_2, q_3, \dots, q_n) \\
&= P(q_1 = x^1)P(q_2 = x^2) \dots P(q_n = x^n) \\
&= p(q_1)p(q_2|q_1) \dots p(q_n|q_{n-1}, q_{n-2}, \dots, q_1)
\end{aligned} \quad (3)
$$

Compare formula (3) with formula (1), we find that if we consider the signal $x_i$ at each time point in formula (3) as the $i$-th word in the sentence, it can exactly represent the conditional probability distribution of each word in the text, which is $p(w_n|w_1, w_2, \dots, w_{n-1})$, and then it can perfectly model the statistical language model of the text. It is because of this commonality that the Markov chain model is very suitable for modeling text and is widely welcomed in the field of natural language processing, especially in the field of automatic text generation.

Generally, in actual situations, the influence of the signal at each moment in the sequence signal on the subsequent signal is limited, that is, there exists a influence domain, and beyond the influence domain, it will not continue to affect the subsequent time signal. Therefore, we assume that for a time-series signal, the value of each time signal is only affected by the first few finite moments. If the value of the signal at each moment is only affected by the signals of the previous m moments, we call it the m-order Markov model and can be expressed as follows:

$$
\begin{aligned}
&P(Q_t = x^t | Q_{t-1} = x^{t-1}, Q_{t-2} = x^{t-2}, \dots, Q_1 = x^1) \\
&= P(Q_t = x^t | Q_{t-1} = x^{t-1}, Q_{t-2} = x^{t-2}, \dots, Q_{t-m} = x^{t-m}), \\
&\text{s. t.} \quad n > t > m
\end{aligned} \quad (4)
$$

When we use the Markov chain model for automatic text generation, we actually hope to use the Markov chain model to obtain a good statistical language model estimate through learning on a large number of text sets. For a big training corpus which contains multiple sentences, we first build a big dictionary $D$ that contains all the words appeared in the training set, that is

$$
D = \{word_{D_1}, word_{D_2}, word_{D_3}, \dots, word_{D_N}\}
$$

where $word_{D_i}$ indicates the $i$-th word in the dictionary $D$ and $N$ is the number of the word. Dictionary $D$ corresponds to the value space $\chi$ described above. As we have mentioned before, each sentence $S$ can be regarded as a sequential signal and the

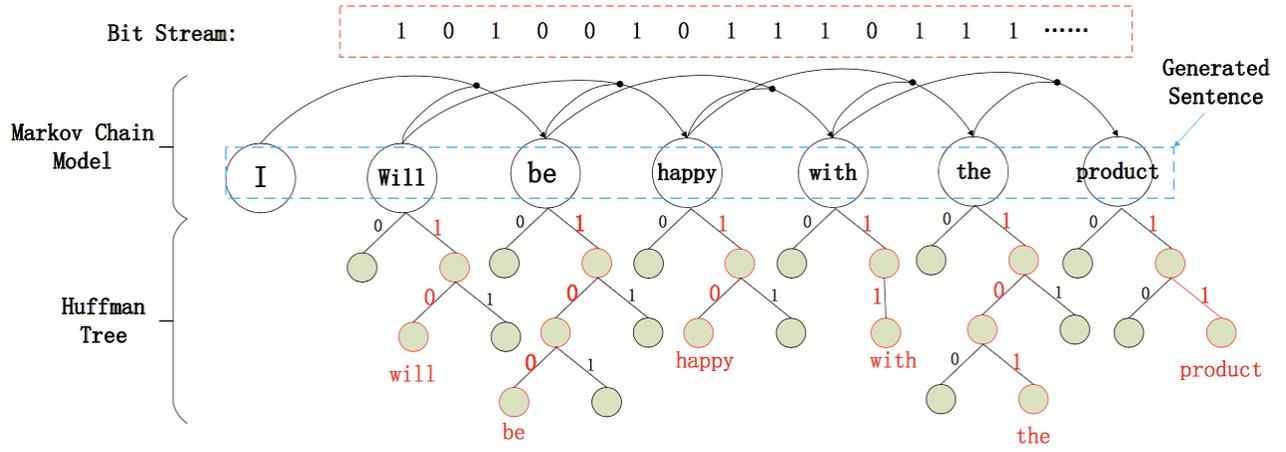

**Figure 1.** A detailed explanation of the proposed model and the information hiding algorithm. The top of the figure is the bit stream that needs to be embedded. The middle part is the Markov chain model (second order) and the generated steganographic sentence. In the text generation process, for each iteration, we construct the corresponding Huffman tree according to the different conditional probability distributions of each word and encode the conditional probability space. And then select the corresponding word according to the secret bits stream, so as to achieve the purpose of hiding the information.

$i$-th word in $S$ can be viewed as the signal at the time point $i$, that is

$$S = \{word_{S_1}, word_{S_2}, word_{S_3}, \ldots, word_{S_L}\},$$
$$\text{s.t.} \quad \forall \; word_{S_i} \in D \tag{5}$$

where $word_{S_i}$ indicates the $i$-th word in sentence $S$ and $L$ is the length of it. In the automatic text generation process, we need to calculate the transition probability of each word. For the Markov chain model, according to the big number theorem, we usually use the frequency of each phrase in the data set to approximate the probability. For example, for a second-order Markov chain model, the calculation formula is as follows:

$$p(word_{S_n} = word_{D_i}|word_{S_{n-2}}, word_{S_{n-1}})$$
$$\approx \frac{count(word_{S_{n-2}}, word_{S_{n-1}}, word_{D_i})}{count(word_{S_{n-2}}, word_{S_{n-1}})},$$
$$\text{s.t.} \sum_{i=1}^{N} p(word_{S_n} = word_{D_i}|word_{S_{n-2}}, word_{S_{n-1}}) = 1, \tag{6}$$

where $count(word_{S_{n-2}}, word_{S_{n-1}}, word_{D_i})$ is the number of occurrences of this phrase $\{word_{S_{n-2}}, word_{S_{n-1}}, word_{D_i}\}$ in the training set. If we don't need to embed information but just generate natural text, we usually choose the word with the highest probability as the output at each iteration.

### 3.2 Information Hiding Algorithm

When we automatically generate texts using the Markov chain model, every time a word is generated, the model calculates the probability distribution $p(Word_{D_i}|Word_{S_1}, Word_{S_2}, \ldots Word_{S_t})$ of the next word according to all the words generated in the previous steps. We encode all the words in the dictionary $D$ based on their conditional probability distribution, and then select the corresponding word according to the secret bit stream, so as to achieve the purpose of hiding the information.

Our thought is mainly based on the fact that when the number of sentences in the sample set for learning is sufficient large, there are actually more than one feasible solution at each time point. After descending the prediction probability of all the words in the dictionary $D$, we can choose the top m sorted words to build the Candidate Pool (CP). To be more specific, suppose we use $c_i$ to represent the $i$-th word in the Candidate Pool, then the CP can be writen as

$$CP = [c_1, c_2, \ldots, c_m].$$

In fact, when we choose a suitable size of the candidate pool, any word $c_i$ in CP selected as the output at that time step is reasonable and will not affect the quality of the generated text, so it becomes a place where information can be embedded. Figure 1 shows the process of generating a complete sentence and embedding secret information using the above model. When we input the keyword "I" at the first time step, Markov chain model will automatically calculate the conditional probability distribution of the next word. By descending the probability of each word in the dictionary $D$, we can select the first eight words to form the candidate pool, then we can get CP = {have, am, will, was, would, bought, got, can}. All of these words can be the output of the next time step and will not make the generated text look weird at all. It is worth noting that each moment when we choose different words, according to the Equation(4), next time step, the probability distribution of the words will be different. After we get the candidate pool, we need to find an effective encoding method to encode the words in it.

In order to make the coding of each word more in line with its conditional probability distribution, we use the Huffman tree to encode the words in the candidate pool. In computer science and information theory, the Huffman code is a particular type of optimal prefix code. The output from Huffman's algorithm can be viewed as a variable length code table for encoding a source symbol. In the encoding process, this method takes fully consideration of the probability distribution of each source symbol in the construction process, and can ensure that the code length required by the symbol with higher coding probability is shorter [31]. In the text generation process, at each moment, we represent each word in the Candidate Pool with each leaf node of the tree, the edges connect each non-leaf node (including the root node) and its two child nodes are then encoded with 0 and 1, respectively, with 0 on the left and 1 on the right, which has been shown in Figure 1.

---

**Algorithm 1**   Information Hiding Algorithm

---

**Input:**
    Secret bit stream: $B = \{0, 0, 1, 0, 1, ..., 0, 1, 0\}$
    Size of Candidate Pool(CPS): $m$
    Keyword list: $A = \{key_1, key_2, ..., key_F\}$
**Output:**
    Multiple generated sentences: $Text = \{S_1, S_2, ..., S_N\}$
    if (not the end of current sentence) then
        Calculate the probability distribution of the next word according to the previously generated words using Markov chain model;
        Descending the prediction probability of all the words and select the top $m$ sorted words to form the Candidate Pool(CP);
        Construct a Huffman tree according to the probability distribution of each word in the CP and encode the tree;
        Read the binary stream, and search from the root of the tree according to the encoding rules until the corresponding leaf node is found and output its corresponding word;
    else
        Random select a keyword $key_i$ in the keyword list $A$ as the start of the next sentence;
    end if
    **Return:** Generated sentences

---

After the words in the Candidate Pool are all encoded, the process of information embedding is to select the corresponding leaf node as the output of the current time according to the binary code stream that needs to be embedded. In order to avoid the condition that two equal sequences of bits produce two equivalent text sentences, we constructed a keyword list. We counted the frequency of the first word of every sentence in the collected texts dataset. After sorting in descending order, we choose the 100 most frequent words to form the keyword list. During the generation process, we will randomly select the words in the keyword list as the beginning of the generated steganographic sentence.

Algorithm details of the proposed information hiding method are shown in Algorithm 1. With this method, we can generate a large number of natural sentences that are syntactically correct and semantically smooth according to the input secret code stream. And then these generated texts can be sent out through the open channel to achieve the purpose of secret information hidden and sent, which has a high concealment.

## 3.3 Information Extraction Algorithm

---

**Algorithm 2**   Information Extraction Algorithm

---

**Input:**
    Multiple generated sentences: $Text = \{S_1, S_2, ..., S_N\}$.
    Size of Candidate Pool(CPS): $m$
**Output:**
    Secret bits stream: $B = \{0, 0, 1, 0, 1, ..., 0, 1, 0\}$.
    for (each sentence $S$ in Text):
        Calculate the probability distribution of the next word according to the previously words using Markov chain model;
        Descending the prediction probability of all the words and select the top $m$ sorted words to form the Candidate Pool(CP);
        Construct a Huffman tree according to the probability distribution of each word in the CP and encode the tree;
        Determine the path from the root node to the leaf node which corresponding to the word at the current moment;
        According to the tree coding rule, ie, the left side of the child node is 0 and the right side is 1, the code stream embedded in the current word is decoded;
        Output the decoded code and append it to $B$;
    end for
    **Return:** Extracted secret bits stream B

---

The process of embedding and extracting secret information is a completely opposite process. After receiving the transmitted steganographic text, the receiver needs to correctly decode the secret information contained therein. The process of information embedding and extraction are basically the same. It is also necessary to calculate the conditional probability distribution of each word at each moment, then construct the same Candidate Pool and use the same coding method to encode the words in the Candidate Pool. It is worth noting that in order to ensure the correct extraction of covert information, both parties need to agree on the use of the same public text data set to construct the Markov chain. Algorithm details of the proposed information extraction method are shown as Algorithm 2.

After receiving the transmitted steganographic text, the receiver first constructs a Markov chain of the same order on the same text data set, then inputs the first word of each sentence as a key into the Markov chain model. At each time point, when the receiver gets the probability distribution of the current word, he firstly sorts all the words in the dictionary in descending order of probability, and selects the top m words to form the Candidate Pool. Then he builds Huffman tree according with the same rules to encode the words in the candidate pool. Finally, according to the actual transmitted word at the current moment, the path of the corresponding leaf node to the root node is determined, so that we can successfully and accurately decode the bits embedded in the current word. By this way, the bits stream embedded in the original texts can be extracted very quickly and without errors.

## 4. EXPERIMENTS AND ANALYSIS

In this section, we designed several experiments to test the proposed model from the perspectives of information concealment and hidden capacity. For concealment, we compared and analyzed the quality of the texts generated at different embedding rates with the training text. For the hidden capacity, we analyzed how much information can be embedded in the generated texts and compared it with some other text steganography algorithms.

### 4.1 Data Preparing

Since we hope our model can automatically imitate and learn the sentences written by humans, we need a large amount of human-written natural texts to train our model and obtain a good enough language model. So we choose three of the most common text datasets as our training sets, and these three datasets are also the most common forms of textual media, which are Twitter [32], movie reviews [33] and News [34].

For Twitter, we chose the sentiment140 dataset published by Alec Go *et al.* [32]. It contains 1,600,000 tweets extracted using the Twitter API. For the movie review dataset, we chose the widely used IMDB dataset published by Maas *et al.* [33]. The texts of the above two datasets are of the social media type. In addition, we also chose a news dataset [34] containing relatively more standard texts to train our model. It contains 143,000 articles from 15 American publications, including the New York Times, Breitbart, CNN and so on. The topics of the dataset are mainly politically related and the published time is mainly between 2016 and July 2017.

Before construct Markov chain model, we need to conduct data pre-processing, which mainly consists of converting all words into lowercase, deleting special symbols, emoticons, web links, and filtering low-frequency words. After pre-processing, the details of the training datasets are shown in Table 1.

Table 1: The details of the training datasets

| Dataset | Twitter [32] | IMDB [33] | News [34] |
|---|---|---|---|
| Average Length | 9.68 | 19.94 | 22.24 |
| Sentence Number | 2,639,290 | 1,283,813 | 1,962,040 |
| Words Number | 25,551,044 | 25,601,794 | 43,626,829 |
| Unique Number | 46,341 | 48,342 | 42,745 |

### 4.2 Imperceptibility Analysisg

The purpose of steganographic system is to hide the existence of information in the carrier to ensure the security of important information. Therefore, the imperceptibility of information is the most important performance evaluation parameter of a steganographic system. Generally speaking, we expect that the steganographic operation will not cause differences in the distribution of carriers in the semantic space. For the steganography methods based on carrier modification, it is possible to ensure that the statistical distribution characteristics of the carrier are unchanged by modifying the region in which the carrier is not sensitive [9]. The model proposed in this paper automatically generated a steganographic carrier based on secret information without giving a carrier in advance. However, it is also necessary to make sure that the generated text carrier should be as consistent as possible with the statistical distribution of the normal carrier, which is actually more challenging.

First, we need to test whether the sentences generated by our model are close enough to the non-steganographic carrier (*i.e.* human written texts) on the statistical language model, otherwise it will be very easy to distinguish. In information theory, *perplexity* is a measurement of how well a probability distribution or probability model predicts a sample. It can be used to compare probability models. In the field of Natural Language Processing, *perplexity* is a standard metric for sentence quality testing [35,36]. It is defined as the average per-word log- probability on the test texts:

$$Perplexity = 2^{-\frac{1}{N}\sum_{i=1}^{N} log p(s_i)}$$
$$= 2^{-\frac{1}{N}\sum_{i=1}^{N} log p_i(w_1,w_2,w_3,...,w_n)}$$
$$= 2^{-\frac{1}{N}\sum_{i=1}^{N} log p_i(w_1)p(w_2|w_1)...p(w_n|w_1,w_2,...,w_{n-1})}, \qquad (7)$$

where $s_i = \{w_1, w_2, w_3, ..., w_n\}$ is the generated sentence, $p(s_i)$ indicates the probability distribution over words in sentence $s_i$, this probability is calculated from the language model of the training texts. $N$ is the total number of generated sentences. By comparing Equation (7) with Equation (1), we find that *perplexity* actually calculates the difference in the statistical distribution of language model between the generated texts and the training texts. The smaller its value is, the more consistent the generated text is with the statistical distribution of the training text.

In order to objectively reflect the performance of our model, we choose two text steganographic methods proposed in [10] and [11] as our baseline. Both of these two methods also use Markov model for steganographic text automatic generation. For each embedding rates, we generated 1, 000 sentences for testing. The mean and standard deviation of the perplexity were tested and the results are shown in Table 2 and Figure 2. Since the number of embedded bits per word (bpw) in our model is uncertain, we calculated the average number of bits per word embedded in the generated text at each CPS.

Based on these results, we can get the following conclusions. Firstly, on each dataset, for each steganography algorithm(except for [10]), as the embedding rate increases, the perplexity will gradually increase, that is, the statistical language distribution difference between the generated text and the training samples will gradually increase. This is because when the number of bits embedded in each word increases, the word selected as the output is increasingly controlled by the embedded bits in each iterative process, and it is increasingly difficult to select the words that match the statistical distribution of the training text best. For [10], they neglect the transition probability of each word in the iterative process, so no matter how many words are selected as candidates, the perplexity of the generated text will remain at a high level. Secondly, the proposed model performance is better than that of the previous two methods in each dataset. This situation is not the cause of generation part, because in the experimental phase, these models use the same Marcov chain model for text generation. The difference in the final results is due to the difference in the coding part. Methods proposed in [10] and [11] do not consider the difference of conditional probability distribution of each word in the coding process. However, for the proposed model, we dynamically code each word based on the conditional probability distribution at each iteration. At different iterations, since the conditional probability distribution changes, it is entirely possible for the same word to have different codings. It is precisely because we fully consider the conditional probability distribution of each word in the coding process, so the text we generate is more in line with the statistical distribution law of the training text.

Table 2: The mean and standard deviation of the perplexity results of different steganography methods at different embedding rates on each dataset.

| Dataset | bpw | Method in [10] | Method in [11] | Ours | bpw(Ours) |
|---|---|---|---|---|---|
| IMDB [32] | 1 | 430.38 ±107.09 | 212.58 ±168.42 | 15.47 ±3.83 | 1.000 |
| | 2 | 432.16 ±110.23 | 269.51 ±155.14 | 20.41 ±5.99 | 1.997 |
| | 3 | 430.61 ±107.19 | 304.71 ±148.39 | 35.04 ±17.63 | 2.940 |
| | 4 | 436.19 ±109.27 | 332.32 ±137.60 | 73.52 ±36.65 | 3.638 |
| | 5 | 433.95 ±110.50 | 348.36 ±131.98 | 137.10 ±64.70 | 3.992 |
| News [33] | 1 | 485.47 ±126.80 | 243.57 ±198.82 | 19.89 ±10.41 | 1.000 |
| | 2 | 487.65 ±133.83 | 302.43 ±186.00 | 27.58 ±17.67 | 1.975 |
| | 3 | 483.03 ±128.49 | 326.62 ±170.20 | 46.14 ±26.93 | 2.879 |
| | 4 | 493.30 ±129.99 | 368.07 ±165.91 | 84.22 ±47.32 | 3.580 |
| | 5 | 485.31 ±132.12 | 382.99 ±151.82 | 151.27 ±78.13 | 3.952 |
| Twitter [34] | 1 | 445.16 ±180.21 | 184.09 ±121.98 | 15.82 ±4.16 | 1.000 |
| | 2 | 445.64 ±166.67 | 257.36 ±135.78 | 22.17 ±8.36 | 1.995 |
| | 3 | 448.52 ±173.66 | 302.66 ±134.94 | 40.56 ±30.60 | 2.942 |
| | 4 | 440.26 ±159.91 | 333.20 ±134.40 | 80.65 ±41.49 | 3.674 |
| | 5 | 440.08 ±166.40 | 349.78 ±124.67 | 143.74 ±72.86 | 4.050 |

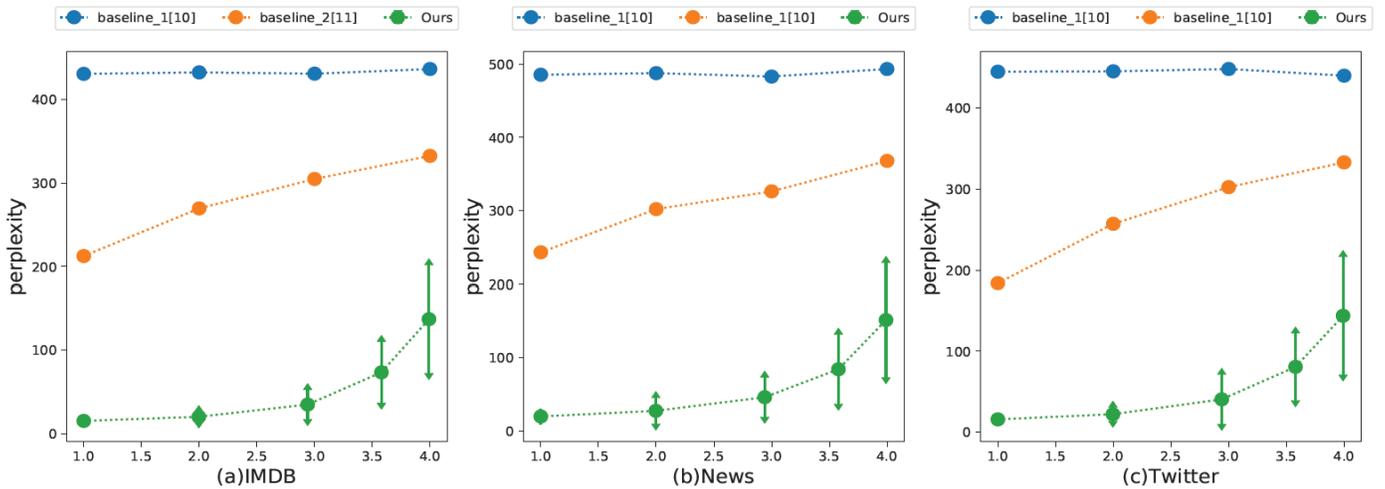

**Figure 2.** The results of different steganography methods at different embedding rates on each dataset.

In addition, we have designed multiple sets of experiments to test the ability of each method to resist steganalysis at high embedding rates (4 bits/word). We implemented the latest text steganographic detection algorithms [37] on steganographic texts generated by each model at an embedding rate of 4 bits/word. The steganalysis method proposed by Samanta *et al.* [37] mainly based on Bayesian Estimation and Correlation Coefficient methodologies. We use several evaluation indicators commonly used in classification tasks to evaluate the steganalysis results, which are accuracy, precision and recall.

- Accuracy calculates the proportion of true results (both true positives and true negatives) among the total number of cases :

$$Acc = \frac{TP+TN}{TP+FN+FP+FN}. \qquad (8)$$

- Precision measures the proportion of positive samples in the classified samples:

$$P = \frac{TP}{TP+FP}. \qquad (9)$$

- Recall measures the proportion of positives that are correctly identified as such:

$$R = \frac{TP}{TP+FN}. \qquad (10)$$

Where TP (True Positive) represents the number of positive samples that are predicted to be positive by the model, FP (False Positive) indicates the number of negative samples predicted to be positive, FN (False Negative) illustrates the number of positive samples predicted to be negative and TN (True Negative) represents the number of negative samples predicted to be negative. Table 3 records the detection results for different steganography algorithms.

Table 3: The steganalysis results of different method.

| Steganalysis [37] | score | Method | | |
|---|---|---|---|---|
| | | [10] | [11] | Ours |
| IMDB [32] | Acc | 0.632 | 0.665 | 0.530 |
| | P | 0.617 | 0.742 | 0.521 |
| | R | 0.700 | 0.670 | 0.535 |
| News [33] | Acc | 0.690 | 0.723 | 0.560 |
| | P | 0.732 | 0.710 | 0.538 |
| | R | 0.641 | 0.719 | 0.620 |
| Twitter [34] | Acc | 0.693 | 0.678 | 0.560 |
| | P | 0.682 | 0.632 | 0.537 |
| | R | 0.655 | 0.850 | 0.605 |

Table 4: some steganographic texts generated by the our model.

| Dataset | bpw | Generated Sentence |
|---|---|---|
| IMDB [32] | 1.000 | i will say this is the best part of the film . |
| | 1.997 | i liked it and this one was a very funny movie . |
| | 2.940 | it's like the film was shot by someone and it was just too stupid plot . |
| News [33] | 1.000 | the government is under investigation . |
| | 1.975 | president trump is a big problem . |
| | 2.879 | he had the chance at an early stage of development . |
| Twitter [34] | 1.000 | i should have a great time . |
| | 1.995 | i dont want 2 go back home from school today . |
| | 2.942 | omg i can't wait for your birthday |

As can be seen from Table 3, on the one hand, when the embedding rate is high (4bit / word), compared with the other two models, our model has the lowest detection rate, indicating that the imperceptibility of our model is better than that of other methods; On the other hand, even if the embedding rate is 4bit/word, the steganographic texts generated by our model are recognized with an accuracy of around 0.5, indicates that the steganographic texts generated by our model are very difficult to identify.

Table 4 shows some steganographic texts generated by our model on different datasets with different embedding rates.

### 4.3 Hidden Capacity Analysis

Embedding Rate(ER) calculates how much information can be embedded in the texts, which is an important index to evaluate the performance of a stenographic algorithm. It forms an opposite relationship with concealment, which usually decreases with increasing embedding rate, as we mentioned earlier, with the number of embedded bits increasing, the quality of generated text decreases. Previous works can hardly guarantee high concealment and large hidden capacity at the same time. In this section, we tested and analyzed the hidden capacity of our model.

The calculation method of embedding rate is to divide the actual number of embedded bits by the number of bits of the whole generated texts. The mathematical expression is as follows:

$$\begin{aligned} ER &= \frac{1}{N}\sum_{i=1}^{N}\frac{(L_i-1)\cdot k}{B(s_i)} \\ &= \frac{1}{N}\sum_{i=1}^{N}\frac{(L_i-1)\cdot k}{8\times\sum_{j=1}^{L_j}m_{i,j}} = \frac{(\bar{L}-1)\times k}{8\times\bar{L}\times\bar{m}} \end{aligned} \qquad (11)$$

where N is the number of generated sentences and $L_i$ is the length of $i$-th sentence. $k$ indicates the number of bits embedded in each word and $B(s_i)$ is the number of bits of the i-th sentence. Since each English letter actually occupies one byte in the computer, ie, 8 bits, the number of bits occupied by each English sentence is $B(S_i) = 8 \times \sum_{j=1}^{L_j} m_{i,j}$ , where $m_{i,j}$ represents the number of letters contained in the $j$-th word of the $i$-th sentence. $\bar{L}$ and $\bar{m}$ represent the average length of each sentence in the generated text and the average number of letters contained in each word. In the actual measurement, we found that the average length of each sentence is 16.95 and the average number of letters contained in each word is 4.79, that is, $\bar{L}$=16.95, $\bar{m}$ = 4.79.

Table 5: The comparison of the embedding rates between the proposed model and the previous algorithms.

| Methods | Embedding Rate (%) |
|---|---|
| Method proposed in [15] | 0.30 |
| Method proposed in [38] | 0.35 |
| Method proposed in [49] | 0.33 |
| Method proposed in [40] | 1.0 |
| Method proposed in [41] | 1.57 |
| Ours (bpw = 3) | 7.34 |

Table 5 shows the comparison of the embedding rates between the proposed model and some previous algorithms which are not based on carrier automatic generation. The line at the bottom is the result of the proposed model when the bits embedded in each word is 3. It can be found from Table 5 that the embedding rate of other types of text steganography algorithms can only be about 1%. However, the embedding rate of the proposed method is much higher than the previous methods, and can reach to 7.34% at 3 bits/word. The previous experiments have shown that when each word is embedded with an average of 3 bits, the proposed model can have a relatively high concealment, while its embedding rate can still achieve 7.34%. If we adjust the size of the candidate pool, the proposed model can even achieve a higher embedding rate. This proves that the proposed model can achieve relatively high concealment and hiding capacity at the same time by adjusting the average bit number of each word embedded.

## 5. CONCLUSION AND FUTURE WORK

The topic that linguistic steganography based on text carrier auto-generation technology is fairly promising as well as challenging. However, due to the high coding degree and less information redundancy in text, it has been an extremely challenging problem to hid information in it for a long time. In this paper, we proposed a steganography method which can automatically generate steganographic text based on Markov chain model and Huffman coding. It can automatically generate fluent text carrier in terms of secret information which need to be embedded. The proposed model can learn from a large number of samples written by people and obtain a good estimate of the statistical language model. We designed several experiments to test the proposed model from several perspectives. Experimental results show that the performance of the proposed model is superior to all the previous related methods in terms of information imperceptibility and information hidden capacity.

## ACKNOWLEDGMENT

This research is supported by the National Natural Science Foundation of China (No.U1536201, No.U1536207 and No.U1636113).